\documentclass[12pt,preprint]{aastex}

\shorttitle{Acetylene toward Cepheus A East}
\shortauthors{Sonnentrucker et al.}

\begin{document}




\title{Detection of Acetylene toward Cepheus A East with {\it Spitzer}}







\author{P. Sonnentrucker\altaffilmark{2}, E. Gonz\'alez-Alfonso\altaffilmark{3} and D. A. Neufeld\altaffilmark{2}}












\altaffiltext{1}{Based on observations with the {\it Spitzer Space Telescope}.}
\altaffiltext{2}{Department of Physics and Astronomy, Johns Hopkins University, 3400 North Charles Street, Baltimore, MD 21218.}
\altaffiltext{3}{Universidad de Alcal\'a de Henares, Departamento de F\'isica, Campus Universitario, E-28871 Alcal\'a de Henares, Madrid, Spain.}







\begin{abstract}
The first map of interstellar acetylene (C$_2$H$_2$) has been
obtained with the infrared spectrograph onboard the {\it Spitzer Space Telescope}. A spectral line map of the $\nu_5$ vibration-rotation band at 13.7 $\mu$m carried out toward the star-forming region Cepheus A East, 
shows that the C$_2$H$_2$ emission peaks in a few localized clumps where 
gas-phase CO$_2$ emission was previously detected with {\it Spitzer}. The 
distribution of excitation temperatures derived from fits to the C$_2$H$_2$ 
line profiles ranges from 50 to 200 K, a range consistent 
with that derived for gaseous CO$_2$ suggesting that both molecules probe 
the  same warm gas component. The C$_2$H$_2$ molecules are 
excited via radiative pumping by 13.7 $\mu$m continuum photons emanating from 
the HW2 protostellar region. We derive column densities ranging from a few 
$\times$ 10$^{13}$ to $\sim$ 7 $\times$ 10$^{14}$ cm$^{-2}$, corresponding 
to C$_2$H$_2$ abundances of 1 $\times$ 10$^{-9}$ to 4 $\times$
10$^{-8}$ with respect to H$_2$. The spatial distribution of the C$_2$H$_2$ emission along with a roughly constant $N$(C$_2$H$_2$)/$N$(CO$_2$) strongly suggest an association with shock activity, most likely the result of the sputtering of acetylene in icy grain mantles.
\end{abstract}




























\section{Introduction}

Acetylene constitutes a key ingredient in the production of large complex hydrocarbon molecules in the dense interstellar medium (Herbst 1995). Because acetylene has no permanent dipole moment, it lacks a rotational spectrum that could be observed at radio wavelengths; observations of interstellar acetylene have therefore been limited to mid-infrared studies of rovibrational bands, carried out from ground-(e.g., Evans, Lacy \& Carr 1991; Carr et al. 1995) and space-based (e.g., Lahuis \& van Dishoeck 2000; Boonman et al. 2003; Lahuis et al. 2007) observatories. Acetylene has been detected in the gas-phase - either in absorption (e.g., Carr et al. 1995; Lahuis \& van Dishoeck 2000) or in emission (e.g. Boonman et al. 2003, this paper) - mostly toward young stellar objects. C$_2$H$_2$ can be used as a tracer of warm (100 K to 1000 K) molecular gas along often complicated sightlines. C$_2$H$_2$ abundance estimates, which were sometimes a few orders of magnitude higher than the predictions of cold gas-phase steady-state chemical models, have led to a better understanding of the role that warm gas-phase chemistry (e.g., Doty et al. 2002; Rodgers \& Charnley 2001) and/or grain mantle processing (e.g., Ruffle \& Herbst 2000) can play in star-forming regions, both locally and in extra-galactic objects (Lahuis et al. 2007).   

In this Letter, we present the first detection of the $\nu_5$ band of acetylene (C$_2$H$_2$) at 13.7 $\mu$m toward the star forming region Cepheus A East using the Infrared Spectrograph (IRS) onboard the {\it Spitzer Space Telescope}. This is the first map of C$_2$H$_2$ obtained toward any interstellar gas cloud. Section 2 describes the observations and data analysis. Sections 3 and 4 compare the spatial distribution of C$_2$H$_2$  to those of gaseous CO$_2$ and H$_2$ $S$(2), and discuss the C$_2$H$_2$-emitting gas in the context of shock chemistry and local outflow activity. The presence of C$_2$H$_2$ on interstellar dust grains will also be discussed in the context of cometary ices composition.

\section{Observations and data analysis}

Spectral maps were obtained with the IRS instrument onboard {\it Spitzer} as 
part of the Guaranteed Time Observer Program 113. 1$'$$\times$1$'$ square
fields were observed with the short-low (SL), short-high (SH), and long-high 
(LH) modules, providing wavelength coverage from 5.2 to 25 
$\mu$m. We obtained continuous spatial sampling by stepping the slit 
perpendicular and parallel to its long axis in steps of one-half its width 
and 4/5 its length, respectively. The data were processed with version 12 
of the pipeline. We used the Spectroscopy Modeling Analysis and Reduction 
Tool (SMART, Higdon et al. 2004), along with locally developed routines 
(Neufeld et al. 2006), to extract wavelength- and flux-calibrated spectra and to 
generate spectral line intensity maps. 

To estimate the uncertainties in the C$_2$H$_2$ line intensities ($\nu_5$ band at 13.7 $\mu$m), 
we first calculated the standard deviation ($\sigma$) around our best fit 
to the local continuum for each point in the map. We then shifted the 
best-fit continuum by $\pm$1 $\sigma$ and generated the corresponding 
new estimates of the line intensities. The difference between the best-fit 
line intensities and the intensities obtained by shifting the best-fit 
continuum model constitutes our $\pm$1 $\sigma$ errors. While extinction due to the long wavelength wing of the silicate feature (9.7 $\mu$m) certainly occurs, corrections to the C$_2$H$_2$ line intensities were not applied. As we will see below, C$_2$H$_2$ arises in a warm component located mostly in front of the cold quiescent gas traced by the silicates (Sonnentrucker et al. 2007, ApJ in press) making a direct estimate of the fraction of dust located in this warm component nearly impossible. Since our study also indicates that the silicates exhibit a homogeneous distribution over the spatial extent of the gaseous C$_2$H$_2$ emission, we are confident that the C$_2$H$_2$ intensity variations we see are not predominantly due to extinction effects.

























\section{Results}

Figure 1 ({\it upper}) shows a summed IRS spectrum in the wavelength range containing acetylene at the spatial location corresponding to one peak in the C$_2$H$_2$ line emission intensity map (HW5/6 in Fig.~2). Note that frequent order mismatches around 14 $\mu$m preclude any reliable detection of the HCN $\nu_2$ band at 14.05 $\mu$m with our data. The striking similarities between the C$_2$H$_2$ maps presented here (Fig.~2) and the gaseous CO$_2$ maps we found previously (Sonnentrucker et al. 2006) strongly suggest that C$_2$H$_2$ arises in the same warm postshock gas component exhibiting gaseous CO$_2$. Because the C$_2$H$_2$ emission lines are weaker by a factor $\sim$10 than those of CO$_2$, we first searched for the spatial positions where C$_2$H$_2$ was detected at a $>$1.5 $\sigma$ level. 38 such individual spatial positions fulfilled this criterion over the mapped region shown in Fig.~2. These are the spectra that we will consider when comparing the C$_2$H$_2$ distribution with that of gas-phase CO$_2$ and H$_2$ $S$(2), a tracer of shock activity (see Fig.~3). 

We constrained the gas temperature in the selected C$_2$H$_2$-emitting region, by comparing synthetic profiles of the $\nu_5$ acetylene band for temperatures ranging from 50 to 900 K with the observed C$_2$H$_2$ $Q$-branch (13.71 $\mu$m) profile. We find that temperatures between 50 and 200 K best fit the 38 selected C$_2$H$_2$ line profiles (Fig.~1, {\it lower}). Note that significantly higher temperatures, such as those found by Lahuis \& van Dishoeck (2000) toward other massive young stellar objects (300-1000 K), are ruled out in Cepheus A East. Most importantly, our moderate 50-200 K range coincides with the temperature range found for gaseous CO$_2$ in that same region (Sonnentrucker et al. 2006) arguing in favor of a common spatial origin for the two species. 

To estimate the column density associated with the C$_2$H$_2$ emission, the dominant excitation mechanism needs to be identified. Accurate collisional rates for vibrational excitation of acetylene are not available to our knowledge. If the collisional rates for C$_2$H$_2$ vibrational excitation were similar to those of CO$_2$ (Boonman et al. 2003), densities higher than $10^8$ cm$^{-3}$ would be required to account for the observed emission. Such a high density would give rise to strong high-lying H$_2$O rotational lines that are, however, not detected (e.g., van den Ancker et al. 2000). Additionally, considering the low temperature in the C$_2$H$_2$-emitting gas component relative to the energy of the $\nu_5=1$ state ($E/k\approx1050$ K), as well as the rather low local hydrogen density inferred over this region ($n_{\rm{H}}$$\sim$ few $\times$ 10$^{3}$ to a few $\times$ 10$^{7}$ cm$^{-3}$; Codella et al. 2005), it is quite unlikely that collisions are the dominant excitation mechanism.  Radiative pumping by 13.7 $\mu$m continuum photons produced by warm dust local to C$_2$H$_2$ is also unlikely to produce the observed emission, since we previously determined that the radiation field in this region is dominated by dust close to the HW2 protostellar region (Sonnentrucker et al. 2006). Therefore, we conclude that radiative pumping by 13.7 continuum photons emanating from the HW2 protostellar region seems the most likely excitation mechanism for the observed acetylene molecules, as we also found for CO$_2$ (Sonnentrucker et al. 2006). 
 
We computed the C$_2$H$_2$ column densities over the mapped region, following the prescription described in Sonnentrucker et al. (2006). Under these assumptions, the derived $N$(C$_2$H$_2$) values are proportional to the acetylene intensities and to the square of the angular separation from the exciting source, HW2 (e.g., Gonz\'alez-Alfonso et al. 2002). Figure 2 displays the intensity map ({\it upper}) and the derived column density map ({\it lower}) for C$_2$H$_2$ toward Cepheus A East. The gray contours show the distribution of NH$_3$ (1,1) a tracer of cold quiescent molecular gas (Torrelles et al. 1993) toward the region fully sampled by {\it Spitzer}. Like gas-phase CO$_2$, the $Q$-branch of gaseous C$_2$H$_2$ mainly traces the walls of the cavity carved by the {\it northeast} component of the outflow originating at HW2 (e.g., G\'omez et al. 1999), an outflow that is apparently responsible for the disruption of the quiescent clouds as seen in the NH$_3$ distribution (e.g., Torrelles et al. 1993; Goetz et al. 1998; van den Ancker 2000). Intensity maxima for C$_2$H$_2$ are detected close to the HW6 radio continuum source, at the NE bridge that joins the Cep A-2 and Cep A-3 cloud cores (Torrelles et al. 1993), and along the western surface of the Cep A-3 cloud. Weaker emission is also found further from the NE bridge. The similarities in both the temperature and the spatial distribution of the C$_2$H$_2$ and CO$_2$ emitting gas add weight to the conjecture that both species arise from the same warm shocked component. 

We finally searched for correlations between the acetylene measurements and those obtained for gas-phase CO$_2$ and H$_2$ $S$(2), a tracer of shock activity, for those 38 positions where C$_2$H$_2$ was detected at a $>$1.5$\sigma$ level. Figure 3 compares the C$_2$H$_2$ intensity with that of gaseous CO$_2$ ({\it upper}), as well as the C$_2$H$_2$ column density with the CO$_2$ column density ({\it middle}) and the H$_2$ $S$(2) intensity ({\it lower}). In all cases, the good correlation found between acetylene, gaseous CO$_2$ and H$_2$ $S$(2) indicates that C$_2$H$_2$ does indeed arise in the warm shocked component also containing gaseous CO$_2$.

\section{Discussion}

Acetylene was previously observed toward low-to-high mass star-forming regions either in absorption (e.g., Lahuis \& van Dishoeck 2000) or in emission (e.g., Boonman et al. 2003) using the {\it Infrared Space Observatory} ({\it ISO}). Excitation temperatures ranging from $\sim$ 10 to 900 K and abundances with respect to H$_2$ ranging from a few $\times$ 10$^{-8}$ to a few $\times$ 10$^{-7}$ were derived. Steady-state models of gas-phase chemistry in cold (10-50K) dense (10$^3$-10$^5$ cm$^{-3}$) molecular clouds predict abundances for acetylene between a few $\times$ 10$^{-10}$ and 1$\times$ 10$^{-8}$ with respect to H$_2$ depending on the role that neutral-neutral destruction reactions may play (Bettens, Lee \& Herbst 1995; Lee et al. 1996). Similar abundances are predicted by models of gas-grain chemistry in quiescent clouds with the highest values obtained only after 10$^6$ years (Ruffle \& Herbst 2000). While such models could account for the observed abundances in the cold gas, they were unable to reproduce the enhancements observed toward much warmer gas components in those objects. Mechanisms such as C$_2$H$_2$ ice sublimation from grain mantles and/or C$_2$H$_2$ enhancements via warm gas-phase chemistry were then invoked (e.g., Carr et al. 1995; Doty et al. 2002; Boonman et al. 2003). 

For the NE outflow region in Cepheus A East, we derive C$_2$H$_2$/H$_2$ abundance ratios in the range 1 $\times$ 10$^{-9}$ to 4 $\times$ 10$^{-8}$, for an assumed H$_2$ column density of 1.5 $\times$ 10$^{22}$ cm$^{-3}$ (G\'omez et al. 1999).  These values are averages along the observed sight-lines.  The highest abundances are localized around and to the south of, the NE position as well as around the positions of the HW5/6 sources (see Fig.~2).  This is precisely where the interaction between the {\it northeast} outflow and the ambient molecular clouds occurs, and it is in these regions that we observe strong H$_2$ $S$(2) emission, a tracer of warm shocked gas. Thus the spatial variation of the C$_2$H$_2$ abundance again strongly suggests an association with shock activity, perhaps as a result of (1) production in the gas phase via high temperature reactions, or (2) grain mantle sputtering.

Models for chemistry in hot cores (Rodgers \& Charnley 2001; Doty et al. 2002) indicate that enhanced abundances of C$_2$H$_2$ ($\sim$ few $\times$ 10$^{-8}$) are expected in warm regions with $T \ge$ 200 K. The good correlations between H$_2$ $S$(2), gaseous CO$_2$ and C$_2$H$_2$ indicate that such high temperatures were reached in the warm gas at the passage of the non-dissociative shock. However, the chemical pathways leading to the production of C$_2$H$_2$ are slow, and enhanced abundances only occur after $\sim$ 10$^4$ years, a time scale much greater than that expected for shock heating of the gas ($\sim$ 300 years; e.g. Kaufman \& Neufeld 1996). Hence, enhanced production of C$_2$H$_2$ by high temperature gas-phase chemistry is unlikely to be predominant in the observed region. Thus the correlations shown in Fig.~3 argue in favor of grain mantle sputtering over gas-phase production as the origin of the C$_2$H$_2$ in Cepheus A East. This is the same production mechanism that we favored for gaseous CO$_2$ (Sonnentrucker et al. 2006). While models for the production of C$_2$H$_2$ in shocks are not available to our knowledge, our results further suggest that both C$_2$H$_2$ and CO$_2$ are released into the gas phase under very similar physical conditions.

The gaseous $\rm C_2H_2/CO_2$ ratio is roughly constant, with a mean value of 0.08 for all sight-lines where we detected acetylene at the 1.5 $\sigma$ level.  If this value reflects the composition of the grain mantle, and given a $N$(CO$_2$)$_{ice}$/$N$(H$_2$O)$_{ice}$ ratio in this source of 0.22 (Sonnentrucker et al. 2007, ApJ in press), then the required $N$(C$_2$H$_2$)$_{ice}$/$N$(H$_2$O)$_{ice}$ ratio is 0.02.  This value is at least a factor 4 larger than those derived toward other star-forming regions (0.1-0.5\%, Evans et al. 1991; Lahuis \& van Dishoeck 2000) and those predicted by theoretical models (0.1-0.5\%, Hasegawa \& Herbst 1993; Ruffle \& Herbst 2000), and at least a factor 2 larger than the gaseous C$_2$H$_2$/H$_2$O ratios obtained in observations of cometary comae (0.1-0.9\%, Brooke et al. 1996; Weaver et al. 1999).  We speculate that these discrepancies might result from (1) the destruction of CO$_2$ by reaction with atomic hydrogen in shocks faster than $\sim$ 30 km s$^{-1}$ (predicted by Charnley \& Kaufman, 2000) and/or (2) a greater efficiency for sputtering of C$_2$H$_2$ in slow shocks.\footnote{Although both these effects would be strongly dependent upon the shock velocity, the relative constancy of the $\rm C_2H_2/CO_2$ would not necessarily require any fine tuning of the shock velocity.  In reality, any sight-line typically samples an ensemble of shocks with a {\it range} of shock velocities, and the constancy of the $\rm C_2H_2/CO_2$ would simply indicate that the admixture of shock velocities varies little from one sight-line to another (e.g. Neufeld et al.\ 2006).}  In either case, the gaseous C$_2$H$_2$/CO$_2$ ratios we observed may exceed the solid C$_2$H$_2$/CO$_2$ ratio, and the $N$(C$_2$H$_2$)$_{ice}$/$N$(H$_2$O)$_{ice}$ ratio could be less than 0.02.  Unfortunately, direct measurements of acetylene ice are not possible, the weak features expected from solid C$_2$H$_2$ being blended with much stronger features of CO and H$_2$O (Boudin et al. 1998).  However, further observations of gaseous C$_2$H$_2$ at a higher signal-to-noise ratio would be very valuable as a probe of any variations in the $\rm C_2H_2/CO_2$ ratio which might provide important clues to the shock physics.

\acknowledgments

This work is based on observations made with the {\it Spitzer Space Telescope}, which is operated by the Jet Propulsion Laboratory under a NASA contract. P.S. and D.A.N acknowledge funding from LTSA program NAG-13114 to the Johns Hopkins University. We are grateful to J.M. Torrelles for providing us with the NH$_3$ maps. We thank the referee for helpful comments.















{\it Facilities:} \facility{Spitzer (IRS)}.

\clearpage























\begin{figure}
\epsscale{.50}
\plotone{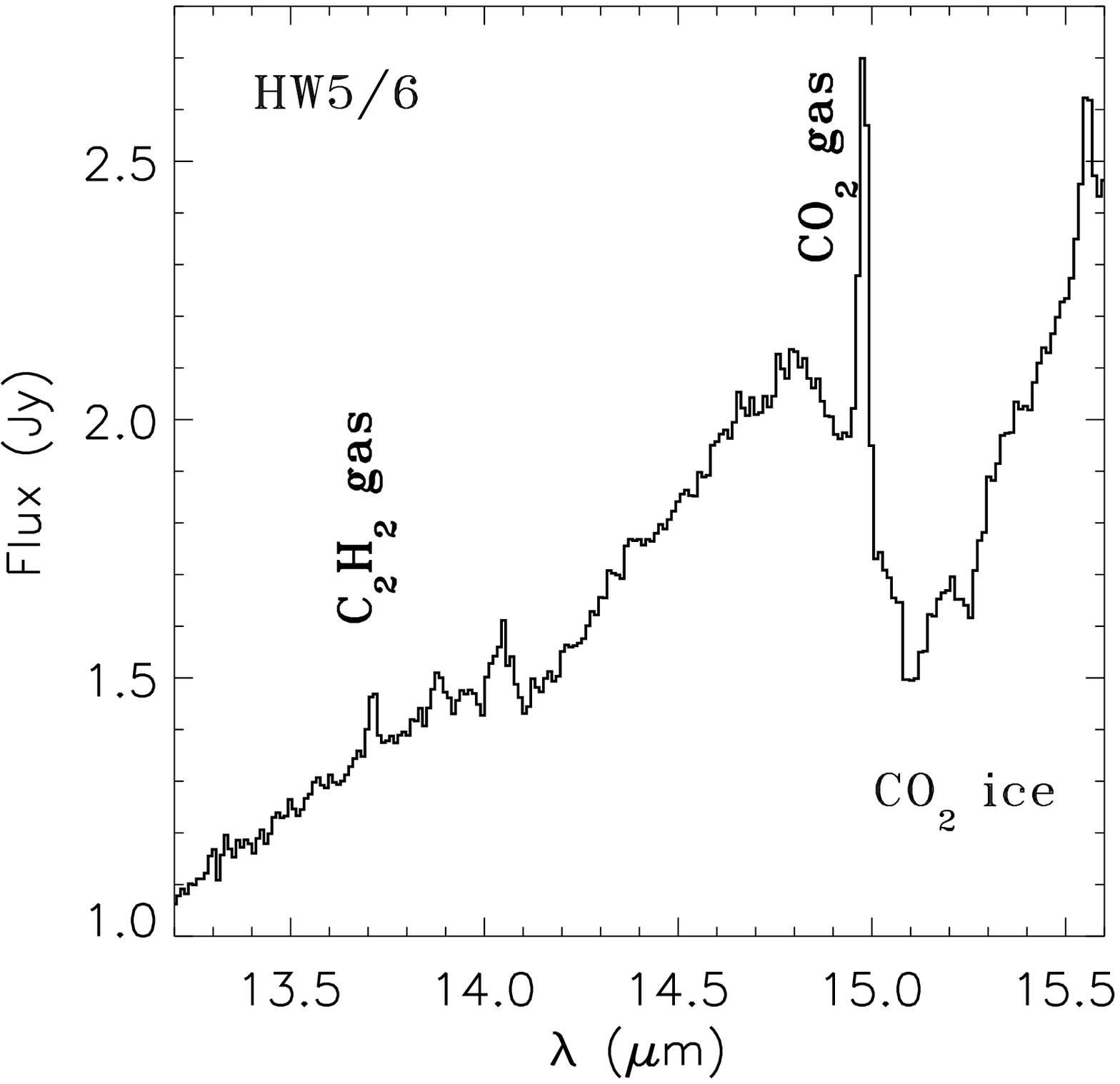}
\plotone{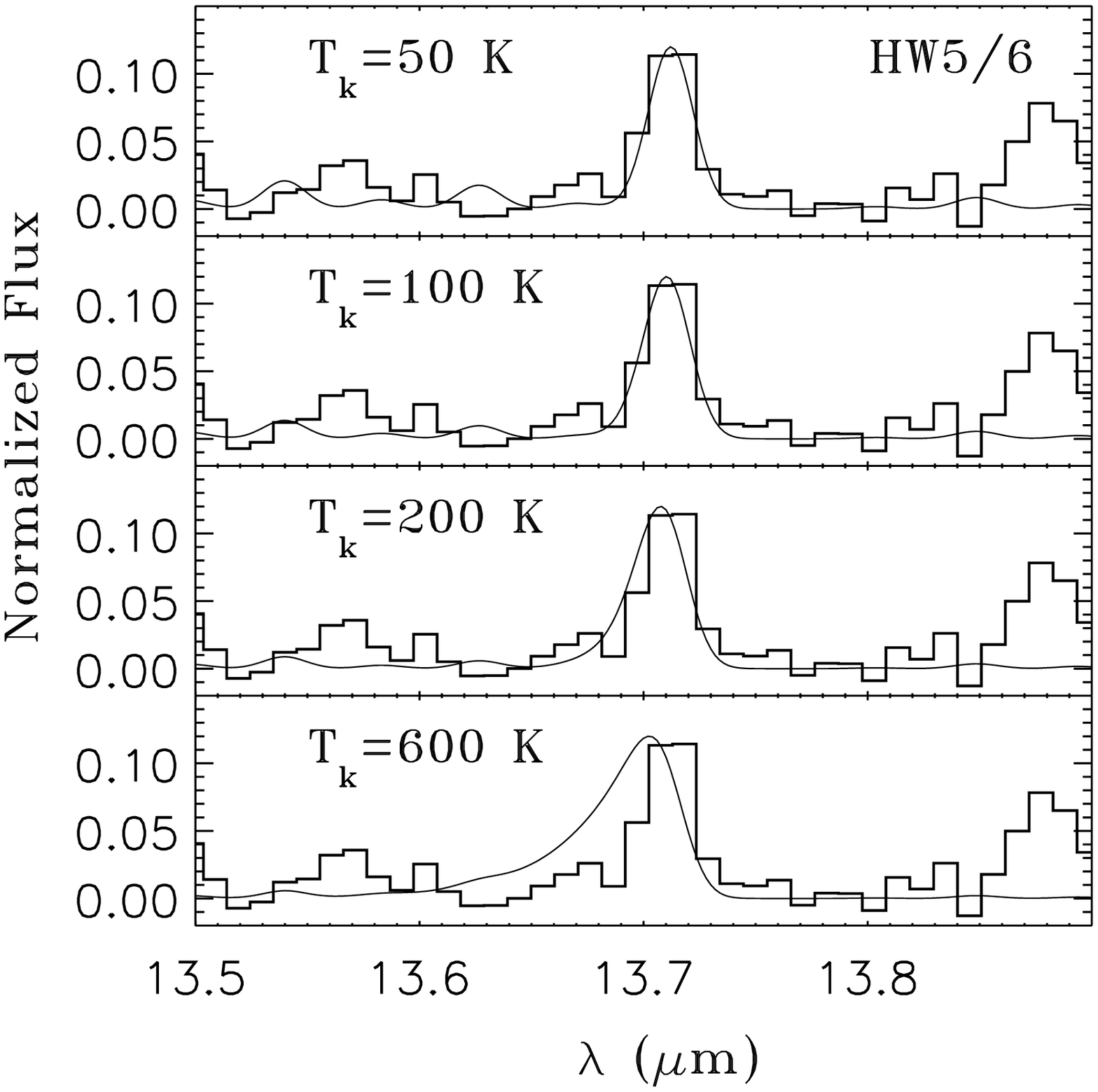}
\caption{\label{fig1} {\it Upper}: Summed IRS spectrum of C$_2$H$_2$ emission toward Cepheus A East. The individual spectra were summed over a $\sim$ 6$''$ $\times$ 8$''$ region centered on the radio-continuum sources HW5/6 positions shown in Figure~2. Note that the emission appearing around 14.05 $\mu$m cannot reliably be attributed to HCN because of a mismatch between the spectral orders containing the C$_2$H$_2$ and CO$_2$ emissions around this wavelength. {\it Lower}: Continuum-subtracted summed spectrum at the HW5/6 positions. Superposed is the calculated C$_2$H$_2$ band spectrum for $T=$ 50, 100, 200, and 600 K. Note the progressive shift of the bandhead and $Q$-branch broadening with increasing temperature.}
\end{figure}






\begin{figure}
\epsscale{.50}
\plotone{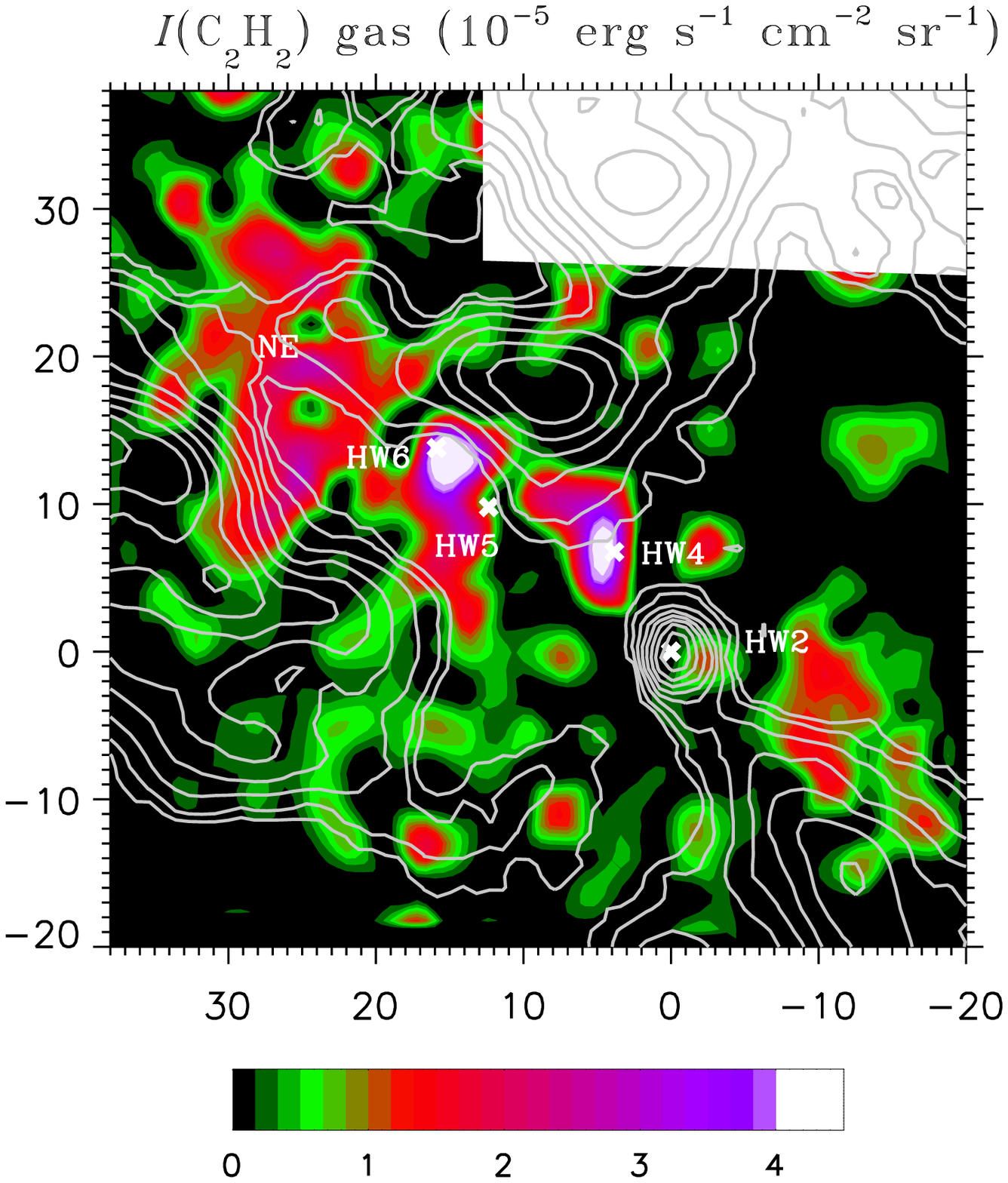}
\plotone{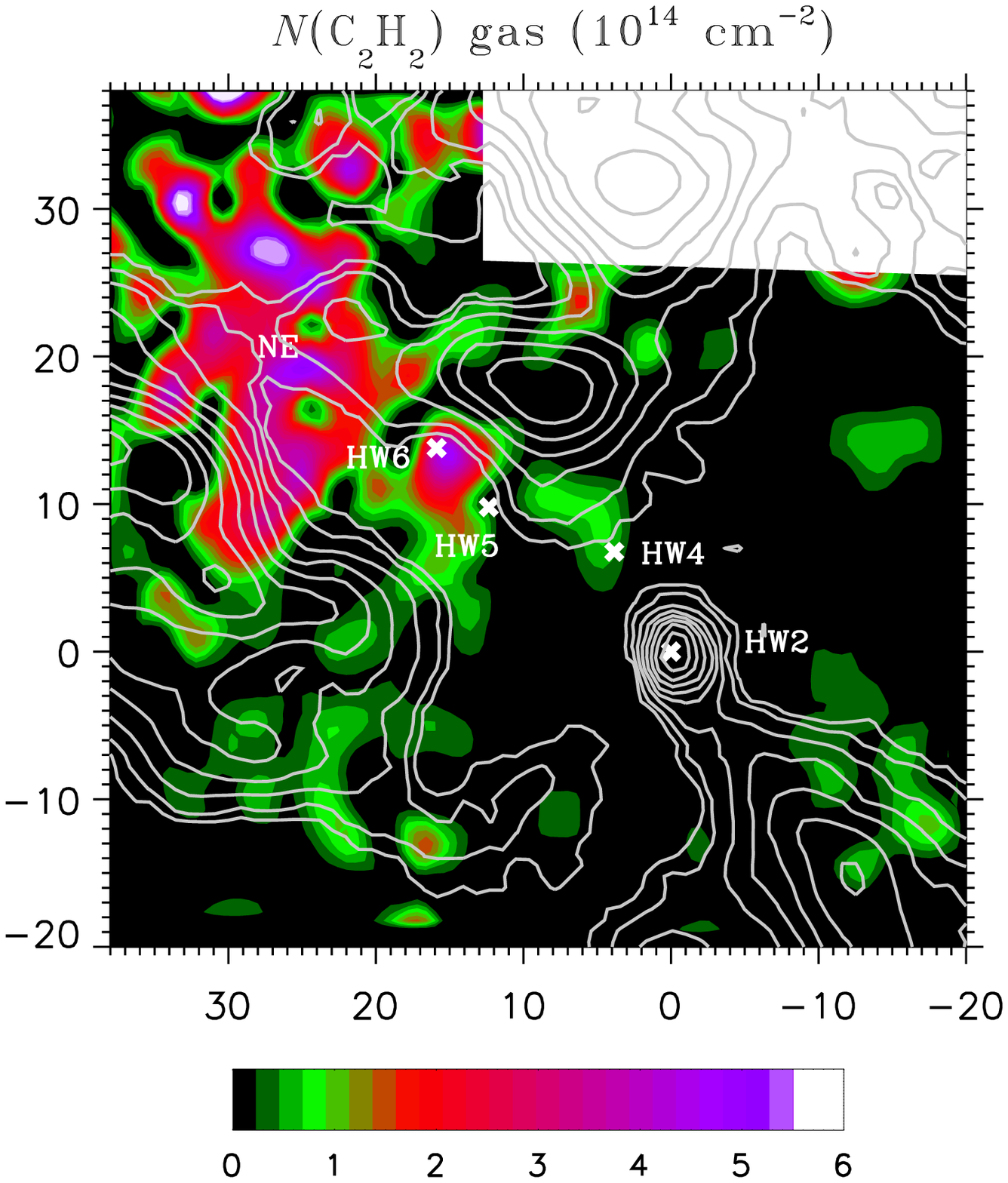}
\caption{\label{fig2} {\it Upper}: Intensity distribution of gas-phase C$_2$H$_2$. {\it Lower}: Column density distribution of gas-phase C$_2$H$_2$. The coordinates are offsets in arcseconds in Right Ascension and Declination with respect to HW2 (J2000.0: $\alpha=$22$^{h}$56$^{m}$17$^s$.9 and $\delta= +$62$^{\circ}$01$'$49$''$; Hughes \& Wouterloot, 1984), the powering source of the outflow oriented {\it northeast} from HW2 (e.g., Goetz et al. 1998). The gray contours show the distribution of NH$_3$ (1,1) (Torrelles et al. 1993), a tracer of quiescent molecular gas. The lowest contours are 10 and 25 in steps of 25 mJy km s$^{-1}$ beam$^{-1}$. The crosses indicate the positions of other known radio-continuum sources (Hughes \& Wouterloot 1984) and NE denotes the edge of the outflow oriented {\it northeast} from HW2.}
\end{figure}

\begin{figure}
\epsscale{.5}
\plotone{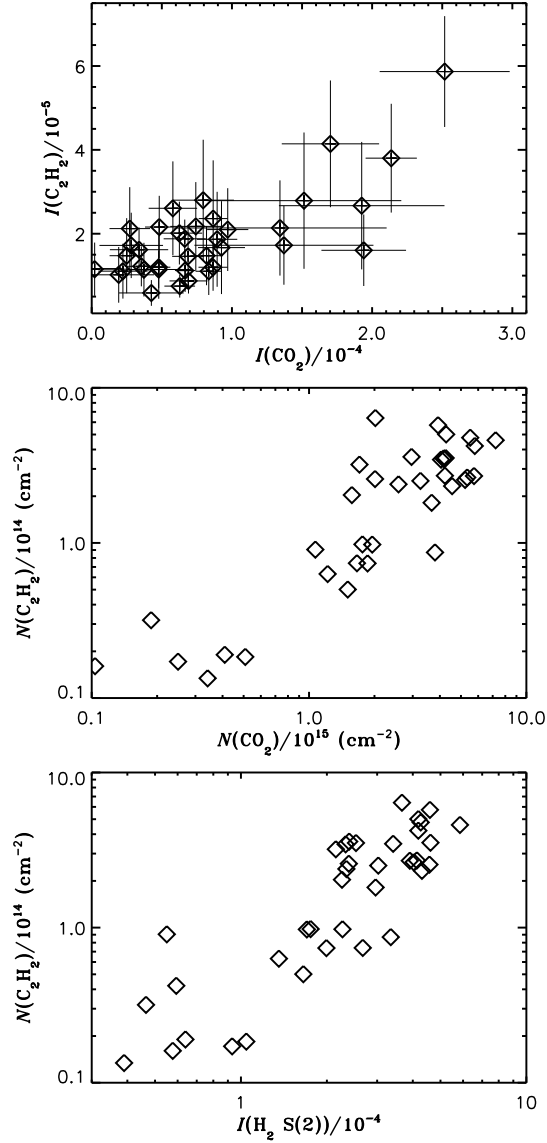}
\caption{\label{fig3} {\it Upper}: Gas-phase C$_2$H$_2$ line intensity compared to the line intensity of gas-phase CO$_2$ with 1$\sigma$ error bars. {\it Middle}: Column density distributions of gas-phase C$_2$H$_2$ and CO$_2$ in logarithmic scale. {\it Lower:} Column density distribution of gas-phase C$_2$H$_2$ compared to the H$_2$ $S$(2) line intensity in logarithmic scale. In all panels, the datapoints correspond to the spatial positions where the C$_2$H$_2$ line intensities were detected at a $>$ 1.5$\sigma$ level. Intensities are in units of ergs s$^{-1}$ cm$^{-2}$ sr$^{-1}$.}
\end{figure}

\end{document}